\documentclass{article}\usepackage[]{graphicx}\usepackage[]{xcolor}
\makeatletter
\def\maxwidth{ %
  \ifdim\Gin@nat@width>\linewidth
    \linewidth
  \else
    \Gin@nat@width
  \fi
}
\makeatother

\definecolor{fgcolor}{rgb}{0.345, 0.345, 0.345}

\usepackage{framed}
\makeatletter
 {\par\unskip\endMakeFramed%
 \at@end@of@kframe}
\makeatother

\definecolor{shadecolor}{rgb}{.97, .97, .97}
\definecolor{messagecolor}{rgb}{0, 0, 0}
\definecolor{warningcolor}{rgb}{1, 0, 1}
\definecolor{errorcolor}{rgb}{1, 0, 0}
\newenvironment{knitrout}{}{} % an empty environment to be redefined in TeX

\usepackage{alltt}
\input{mytheme.sty}
\usepackage{tikz}
\usepackage{booktabs}
\usetikzlibrary{arrows.meta, shapes, arrows, calc,fit, positioning,shapes}
\tikzstyle{block} = [rectangle, draw,
    text width=5em, text centered, rounded corners, node distance=3cm]
\tikzstyle{block2} = [rectangle, draw,
    text width=7em, text centered, rounded corners, node distance=4cm]

\tikzstyle{block3} = [rectangle, draw,
    text width=20em, text centered, rounded corners, node distance=2cm]

    \tikzstyle{block4} = [rectangle, draw = none,
    text width=20em, text centered, rounded corners, node distance=1.5cm]

\tikzstyle{plain2} = [draw = none, fill = none,
text width=5.5em, minimum height = 1cm]
\tikzstyle{cloud} = [draw, ellipse, node distance=5cm,minimum height=5em]
\tikzstyle{line} = [draw, -latex]%-stealth]
\tikzstyle{plain} = [draw=none, fill=none]

\title{\bf{The replication of equivalence studies}}
\author{Charlotte Micheloud and Leonhard Held\\ 
  Epidemiology, Biostatistics
  and Prevention Institute (EBPI)\\ 
  and Center for Reproducible Science (CRS) \\
  University of Zurich\\ Hirschengraben 84,
  8001 Zurich, Switzerland\\ Email: \texttt{charlotte.micheloud@uzh.ch}}

\IfFileExists{upquote.sty}{\usepackage{upquote}}{}
\begin{document}

% some settings

\maketitle

\vspace{-.5cm}
\begin{center}
\begin{minipage}{12cm}
\paragraph{Abstract} 
Replication studies are increasingly conducted to assess the credibility of scientific findings. 
Most of these replication attempts target studies with a superiority 
design, but there is a lack 
of methodology regarding the analysis of replication studies with alternative 
types of designs, such as equivalence.
In order to fill this gap, we propose two approaches, the two-trials rule and the 
sceptical TOST procedure, adapted from methods used in superiority settings.
Both methods have the same overall Type-I error rate, but the sceptical TOST procedure allows 
replication success even for non-significant original or replication studies. 
This leads to a larger project power and other differences in relevant operating characteristics.  
Both methods can be used for sample size calculation of the 
replication study, based on the results from the original one. 
The two methods are applied to data
from the Reproducibility Project: Cancer Biology. 
\end{minipage}
\end{center}

\section{Introduction}\label{sec:intro}
Replicability of scientific findings is the gold standard to assess their 
credibility. Recent years have witnessed an increased interest in large-scale 
replication projects, aiming to reproduce the results found in an original 
study in one or several replication studies. Various empirical domains
of science are involved in these replication efforts: 
psychology \citep{open2015}, social sciences \citep{Camerer2016}, 
economics \citep{Camerer2018} and 
more recently
cancer biology \citep{Errington2021}, among others.

In the majority of these endeavors, there is evidence
against a point null hypothesis in the original study, 
\ie the original effect estimate is significant or borderline significant.
A replication
study is then conducted to confirm the initial result and a
significant replication effect estimate in the same direction is
generally interpreted as a successful replication. Requiring two
statistically significant studies is analogous to the two-trials rule
in drug development \citep{FDA1998}.  But the two-trials rule is not
the only criterion used in the assessment of replication success;  
many replication projects also consider other measures such
as compatibility of the effect estimates from both studies and
meta-analysis of these estimates.  Furthermore,
there is a growing body of literature on the design and analysis of
such replication studies, and new methodology is emerging (\eg in
\citet{Anderson2017, Bonett2020, Held2020, Hedges2021,
MicheloudHeld2021, HeldMichPaw2021, Micheloud2023}).

However, it might also happen that original studies with a `null 
result', \ie with a $p$-value considerably larger than the significance level,
are 
selected for replication. 
This was for example the case for 15 out of
112 effects in the Reproducibility Project Cancer Biology 
\citep[RPCB,][]{Errington2021} and for
3 out of 100 studies in the Reproducibility Project Psychology 
\citep[RPP,][]{open2015}. One criterion for replication success 
of such effects was a non-significant replication effect estimate. 
However, interpreting a non-significant effect as null can be 
misleading, as the apparent null effect could in reality be caused 
by a low sample size, which, if increased, would 
render the same effect statistically significant. This issue 
has also been pointed out by \citet{Pawel2023}.
In fact, if the aim of the original (and subsequently the replication) study 
is to show that an effect is null,
then another type of design needs to be used in both studies:
an equivalence design. 
Equivalence studies are conducted to show the opposite of superiority studies,
namely that an effect $\theta$ is sufficiently close to 0 that it can
be considered negligible. 
Because it is impossible to accept a point 
hypothesis, an interval of equivalence $[-\delta;\delta]$ needs to be 
specified \citep{Serlin1985}.
In clinical trials regulations, 
the International Conference on Harmonisation (ICH) E9 \citep[p. 18]{ICH1999}
states that the margin $\delta$ is `the largest difference that can be judged 
as being clinically acceptable and should be smaller than differences 
observed in superiority trials of the active comparator'. 
The composite hypotheses $H_0: \theta \notin [-\delta, \delta]$ and 
$H_1: \theta \in [-\delta, \delta]$
can be decomposed into two one-sided hypothesis pairs, namely
\begin{equation}\label{eq:set2}
\mbox{$H_0^{+}$ $: $ } \theta \geq \delta \text{ vs. } \mbox{$H_1^{+}$ $:$ } 
\theta < \delta
\end{equation}
and
\begin{equation}\label{eq:set1}
\mbox{$H_0^{-}$ $: $ } \theta \leq -\delta \text{ vs. } \mbox{$H_1^{-}$ $:$ }
\theta > -\delta \, .
\end{equation}

This decomposition is the basis of a widely used method in the assessment of equivalence 
known as the 
`two one-sided tests (TOST)' procedure \citep{Schuirmann1987}. 
Equivalence of the unknown effect $\theta$ is declared if and only 
if both null hypotheses in \eqref{eq:set2} and \eqref{eq:set1} 
are rejected in favor of the alternative
at the one-sided nominal level of significance $\alpha$.
An alternative, equivalent approach is to calculate a $(1 - 2 \alpha)$\%
confidence interval for $\theta$ and conclude equivalence if it 
falls entirely within the interval $[- \delta; \delta]$.
The Type-I error rate of the TOST procedure, \ie the probability 
to incorrectly declare equivalence, depends on the value of the true 
effect $\abs{\theta} \geq \delta$, 
and is at most $\alpha$ \citep[Section 11.5.2]{mat2006}. 

% One 
% approach proposed by \citet{Pawel2023} to assess equivalence
% in the replication setting is to apply the TOST procedure 
% to both the original and the replication study, thus adapting
% the standard two-trials rule in superiority studies.
% However, the theoretical 
% properties of this method have not been studied yet and sample size calculation
% for the replication study is not discussed.
%

Equivalence designs have been used for many years now, especially in clinical 
trials and pharmacokinetics, and the methodology has been extensively discussed.
Similarly, replication studies are very popular nowadays, and numerous 
methods for their design and analysis in the superiority setting 
have been proposed.
However,
there has been little interest so far in combining the two topics:
the original study has an equivalence design and the replication aims to confirm
this equivalence.
The primary aim of this paper is hence to fill this 
gap and to present methodology to analyze and design equivalence
replication studies. In particular, we investigate in detail the theoretical 
properties and the error rates
of applying the TOST procedure twice (simply called the two-trials rule here)
as proposed in \citet{Pawel2023}. Furthermore, we develop the 
`sceptical TOST procedure', 
which stems from the sceptical $p$-value \citep{Held2020}, a reverse-Bayes method
to assess replicability in superiority studies.
In \citet{Micheloud2023}, a recalibration gives rise to the \emph{controlled} sceptical $p$-value, 
ensuring the same Type-I error rate as the two-trials rule when the true effect is null 
in both studies.
The controlled sceptical $p$-value has a number of advantages over the two-trials 
rule in the superiority setting, which are summarized in Section~\ref{sec:scTOST}.

The paper is structured as follows: 
The two-trials rule applied to equivalence studies and the
sceptical TOST procedure are presented
and their properties 
are described in Section~\ref{sec:methods}.
The two methods are then further compared in Section~\ref{sec:power} with a focus on 
power and Type-I error rate. 
Sample size calculations
for an equivalence replication study using the
information provided by the original result are described in
Section~\ref{sec:design}.
Finally, some discussion is provided 
in Section~\ref{sec:discussion}.
Two study pairs from \citet{Errington2021}
are presented in Section~\ref{sec:runningex}, and 
will be used as running examples.

% --------------------

\subsection{Statistical framework}

In our framework the original and replication
effect estimates $\hat\theta_o$ and $\hat\theta_r$ are assumed to be normally 
distributed around the unknown effects $\theta_o$ and $\theta_r$, respectively,
with standard error 
$\sigma_o$ and $\sigma_r$.
The squared standard errors can generally be expressed as
$\sigma_o^2 = \kappa^2/n_o$ and $\sigma_r^2 = \kappa^2/n_r$ 
where $\kappa^2$ is some unit variance and  
$n_o$ and $n_r$ the original and replication sample sizes, respectively. 
The variance ratio $c = \sigma_o^2/\sigma_r^2$ then reduces to 
the relative sample size
$c = n_r/n_o$. This also holds in the balanced two-sample case where 
$\sigma_i^2 = 2\, \kappa^2/n_i$, $i \in \{o, r\}$, with $n_i$
the sample size per group.
The equivalence interval $[- \delta; \delta]$ is 
assumed to be symmetric around $0$
and identical for the original 
and the replication study. 
Furthermore, the one-sided significance level 
$\alpha$ is set to $5$\%, and $z_\alpha = \Phi^{-1}(1 - \alpha)$ 
is the $(1-\alpha)$-quantile of the standard normal distribution 
function, with $\Phi(\cdot)$ its cumulative distribution function.
Without loss of generality, the original effect estimate 
$\hat\theta_o$ is assumed to be positive and smaller than the margin 
$\delta$.

A focus of this paper is Type-I error (T1E) control, so
definitions of the relevant null hypotheses are required. The study-specific null hypotheses
are denoted by $H_0^{i}$, $i \in \{o, r\}$, and depend on the design type:
$H_0^{i}: \theta_i = 0$ for superiority and 
$H_0^{i}: \theta_i \notin [- \delta; \delta]$ for equivalence.
We follow \citet{Heller2014} and \citet{Micheloud2023} and differentiate between the intersection
null hypothesis
\begin{equation*}\label{eq:intersection}
H_0^{o} \cap H_0^{r} \, 
\end{equation*}
and the union null hypothesis
\begin{equation*}\label{eq:intersection}
H_0^{o} \cup H_0^{r} \, .
\end{equation*}
% .
% 
The probability of a false claim of replication success 
associated with these two hypothesis types are 
the \emph{overall} and the \emph{partial} T1E rate, respectively.
The risk of a false claim of replication success if the 
replication study sample size is properly selected based on the original 
study result is referred to as the \emph{conditional T1E rate}.

% --------------------

\subsection{Examples}\label{sec:runningex}

In order to illustrate our methodology, we will use two study pairs 
from \citet{Errington2021} (dataset available at \url{https://osf.io/39s7j}). 
In this project, a nested structure was employed, where each paper selected for 
replication contained several experiments. These experiments, in turn, 
encompassed more than one effect of interest, and finally, certain effects were
replicated in multiple internal replications.
Original effects with a non-significant $p$-value 
were termed `null' by \citet{Errington2021},
and replication success was achieved with a non-significant 
replication study. As discussed before,
this approach is problematic as
non-significance in both studies does not guarantee that there is 
evidence for no effect \citep{Pawel2023}.  
% We propose two alternative replication success
% criteria which do not suffer from this problem in Section~\ref{sec:methods}.
%
% 
% We now present two study pairs from~\citet{Errington2021} 
% They will be used as running examples to illustrate our 
% methodology and to highlight why absence of evidence should 
% not be considered as evidence of absence. 
% and one criterion for replication success which they used
% for such original effects is the `non-significance' criterion, namely a 
% non-significant replication effect was required as well. 
%
%
In the original study from \citet{Goetz2011} \citep[Paper \#20, Experiment \#1, 
Effect \#1 in][]{Errington2021}, there was no evidence 
for a treatment effect
(two-sided $p = 0.34$).
The replication team also failed to find a significant
effect (Internal replication \#1, $p = 0.83$), 
hence replication success was declared.
%
% As the original non-significant result was `replicated',
% replication success 
% was declared with the `non-significance' criterion in \citet{Errington2021}.
In contrast, the non-significant result in the original study by 
\citet{Lin2012} (Paper \#48, Experiment \#2, 
Effect \#4)
was not successfully replicated,
as a significant effect estimate was detected in the 
replication study (Internal replication \#2, two-sided $p  < 0.0001$). 

The dataset
gives the results as
correlation coefficients $r$. We transform them with Fisher 
$z$-transformation $z = \arctanh(r)$ to achieve a normal distribution.
The standard error of $z$ 
is a function of the \emph{effective} sample size 
$n - 3$ only: $\se(z) = 1/\sqrt{n - 3}$.
\begin{figure}[!h]
  \centering
\begin{knitrout}
\definecolor{shadecolor}{rgb}{0.969, 0.969, 0.969}\color{fgcolor}
\includegraphics[width=\maxwidth]{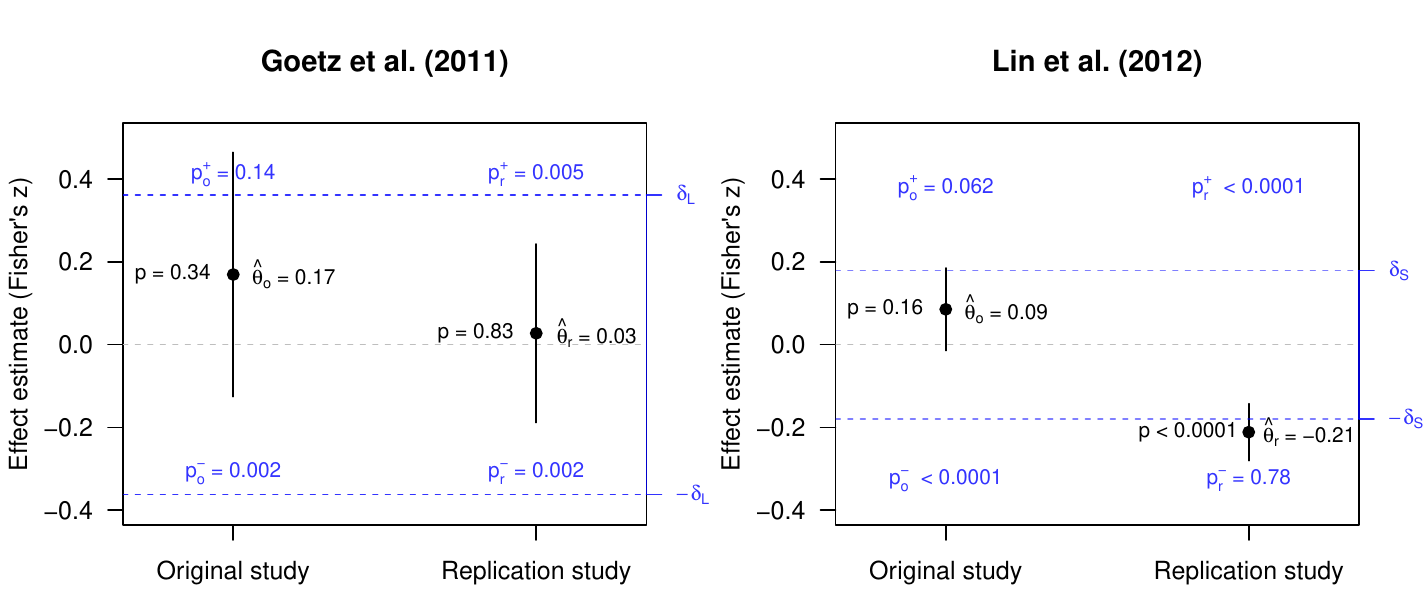} 
\end{knitrout}
  \caption{Original study by \citet{Goetz2011} and its replication by 
  \citet{Sheen2019} and original study by \citet{Lin2012}
  and its replication \citep{Blum2015}.
  Shown are the effect estimates on Fisher's $z$-scale 
  with 90\% original and replication confidence intervals: 
  $[-0.13; 0.47]$ and $[-0.19; 
  0.24]$, respectively, in the first study pair, 
  and $[-0.01; 0.19]$, respectively 
$[-0.28; -0.14]$ in the 
second study pair. The $p$-values in black are two-sided $p$-values 
from the superiority tests, while the $p$-values in color are one-sided
$p$-values from the TOST procedure as explained in Section~\ref{sec:2tr}.
  The strict $\delta_S = 0.18$ and the liberal 
  $\delta_L = 0.36$ margins are indicated 
  with colors.
The original and replication effect estimates in the 
  second study pair have been re-oriented to make
  $\hat\theta_o$ positive.}
  \label{fig:Goetz2011}
\end{figure}
Figure~\ref{fig:Goetz2011} shows the original and replication effect estimates 
(on Fisher's $z$-scale) with their 90\%-CI, so $\alpha = 5$\%.
The standard errors are 
$\sigma_o = 0.18$ and 
$\sigma_r = 0.13$ and so 
$c = \sigma_o^2/\sigma_r^2 = 1.9$ in the 
first study pair, and  $\sigma_o = 0.06$ and 
$\sigma_r = 0.04$ and so 
$c = \sigma_o^2/\sigma_r^2 = 2.1$ in 
the second study pair.
Note that the variance ratio $c$ is equal to the ratio (replication to 
original) of the {effective}
sample sizes. These study pairs will serve as running examples to illustrate our 
methodology and to highlight why
non-significant effects should not be misinterpreted as bearing evidence for
no effect.

In equivalence designs, the margin should ideally be chosen by experts in the 
field before the first study is conducted, but post-hoc margin specification
is also possible \citep{Campbell2021}. As no margin was pre-specified in 
\citet{Goetz2011} and \citet{Lin2012}, we have to select the margins.
\citet[Table 1.1]{Wellek2010} defines a strict margin to be $d_S = 0.36$ 
and a liberal margin $d_L = 0.74$ on Cohen's $d$ scale. This translates to 
$\strict = 0.18$ and $\lib = 0.36$
on Fisher's $z$-scale, respectively \citep{Ruscio2008}.
For illustration purposes, we chose to use the liberal margin in the first example,
and the strict margin in the second.
% We do not use the strict margin $\delta_s = round(strict_fis, 2)$ in the first example 
% as it is almost equal to the 
% effect estimate $\hat\theta_o = round(study$fiso[1], 2)$, so there is 
% clearly no equivalence possible.
%
These margins might seem large, however,  a systematic review of the margins 
in non-inferiority and equivalence clinical trials revealed 
that 50\% of the trials had a margin larger than $d = 0.5$, 
and 5\%  even had a margin larger than $d = 1$ \citep{Lange2005}.
%----------------------------

% ---------------------------------------------------------------------------

\section{Methods}\label{sec:methods}
In this section, two approaches are presented: the two-trials rule
for equivalence studies (Section~\ref{sec:2tr})
and the sceptical TOST procedure (Section~\ref{sec:scTOST}), 
summarized in Figure~\ref{fig:2tr_rs}.
Some properties of both 
methods are then described (Section~\ref{sec:upperBounds})
and an application is given in Section~\ref{sec:appli}.
\subsection{Two-trials rule}\label{sec:2tr}

The two-trials rule applied to an equivalence design
amounts to applying the TOST procedure twice --
once for the original and
once for the replication study \citep{Pawel2023}.
This results in four one-sided tests with corresponding $z$-values
defined as
\begin{eqnarray*}
 z_o^{+} = (\hat\theta_o - \delta)/\sigma_o \, , &   
z_r^{+} = (\hat\theta_r - \delta)/\sigma_r  &\, \mbox{ for $H_o^{+}$ in~\eqref{eq:set2},  and } \, \\
z_o^{-} = (\hat\theta_o + \delta)/\sigma_o \, , &  
z_r^{-} = (\hat\theta_r + \delta)/\sigma_r & \mbox{ for $H_o^{-}$ in~\eqref{eq:set1}} \, . 
\end{eqnarray*}
The $z$-values $z_i^{+}$ and $z_i^{-}$, 
$i \in \{o, r\}$, are related to each other via
\begin{eqnarray}\label{eq:zo2zo}
z_i^{-} = \left(\frac{f_i + 1}{f_i - 1}\right) z_i^{+} \, ,
\end{eqnarray}
with
\begin{equation*}\label{eq:f}
f_i = \hat\theta_i/\delta \, .
\end{equation*}
% indicating how close the effect estimate $\hat\theta_i$ is to the margin $\delta$. 
%
The $z$-values are then transformed to $p$-values using
$p_i^{+} = \Phi(z_i^{+})$ and $p_i^{-} = 1 - \Phi(z_i^{-})$, $i \in \{o, r\}$.
The two-trials rule is fulfilled if 
$p^{\mbox{\tiny max}} = \max\{\pomax, \prmax \} < \alpha$, with
$\pomax = \max\{p_o^{+}, p_o^{-}\}$ and $\prmax = \max\{p_r^{+}, p_r^{-}\}$, see Figure~\ref{fig:2tr_rs}.
%
% \hl{The procedure is illustrated in Figure~\ref{fig:2tr_rs}.}
Note that with a positive original effect estimate $\hat\theta_o$, 
$p_o^{-}$ is always smaller than $p_o^{+}$, and so 
$\pomax = p_o^{+}$.

% ----------------------------------------------------------------
% success region

% ---------------------------------------------------------------------- %
\subsection{The sceptical TOST procedure}\label{sec:scTOST}

The sceptical TOST procedure is an adaptation of the sceptical $p$-value 
which has so far only been developed for superiority studies. 
The theory behind the sceptical $p$-value
is first briefly summarized, and our adaptation to equivalence studies
is then presented.

\paragraph{Superiority studies}
\citet{Held2020} proposed the sceptical 
$p$-value, a new approach to declare replication success in
%quantitative measure of replication success
superiority studies
based on two steps: the analysis of 
credibility \citep{Matthews2001, Matthews2001b} and a prior-predictive check for conflict \citep{Box1980}.
% will need to add that the original study is significant
In a nutshell, a significant original result at level $\alpha$
is combined with a sceptical prior centered around zero, and with 
variance chosen such that the significant result is no longer convincing (lower limit 
of the ($1 - \alpha) \times 100$\% posterior credible interval is
exactly equal to $0$).
If there is sufficient conflict 
between the sceptical prior and the replication result,
the sceptical prior is deemed unrealistic and replication success is declared. 
% 
% The second step consists of measuring the conflict between the sceptical 
% prior and the replication data. 
The quantitative measure of replication success resulting from this procedure 
is the sceptical $p$-value, which only depends on the study-specific $p$-values $p_o$ and $p_r$, and the relative sample size $c$. Replication success is achieved if the 
sceptical $p$-value is smaller than the level $\alpha$.
% say that the sceptical $p$-value measures the conflict, hence the 
% degree of replication success

\begin{figure}[!h]
  \centering
\begin{knitrout}
\definecolor{shadecolor}{rgb}{0.969, 0.969, 0.969}\color{fgcolor}
\includegraphics[width=\maxwidth]{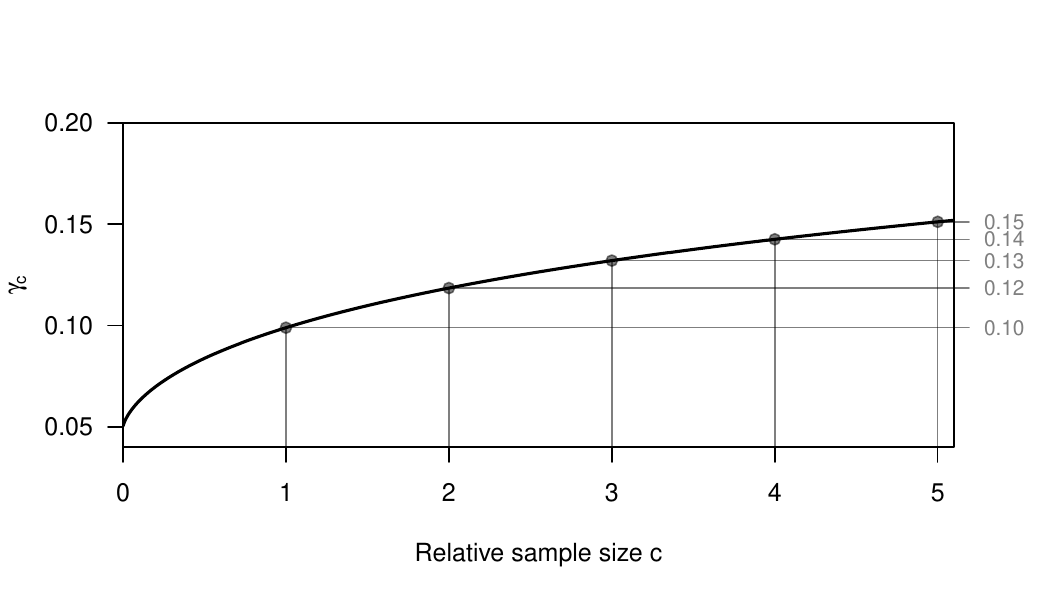} 
\end{knitrout}
\caption{Bound $\gamma_c$ on the partial T1E rate for superiority 
studies as a function of the
relative sample size $c$ for $\alpha = 0.05$.}
\label{fig:repSuccess}
\end{figure}

The original formulation was recently recalibrated by \citet{Micheloud2023}
to give rise to the 
\emph{controlled sceptical $p$-value} $p_S$
% which can be thresholded at $\alpha$ as an ordinary $p$-value, 
% so $p_S < \alpha$ is required for replication success.
% The controlled sceptical $p$-value
which ensures that the overall T1E rate is 
exactly controlled at $\alpha^2$. 
% the same as with the two-trials rule in superiority studies.
The theory in \citet{Micheloud2023} has also shown that
the corresponding partial T1E rate 
% of the sceptical $p$-value 
is bounded by
$\gamma_c  > \alpha$, see Figure~\ref{fig:repSuccess}. 
The controlled sceptical $p$-value (simply called the sceptical $p$-value from 
now on) hence does not impose a strict dichotomization 
at $\alpha$ for the original and replication $p$-values as the 
two-trials rule,
and success is even possible for non-significant original or replication studies 
as long as the two
$p$-values are smaller than $\gamma_c$.
Attractive properties of the approach include an increased project 
power to detect existing effects over both studies in combination and the 
possibility to calculate the replication sample size even for 
non-significant original studies. Moreover, the conditional T1E rate is 
sufficiently bounded \citep[Section 3.4]{Micheloud2023}.

\paragraph{Equivalence studies}

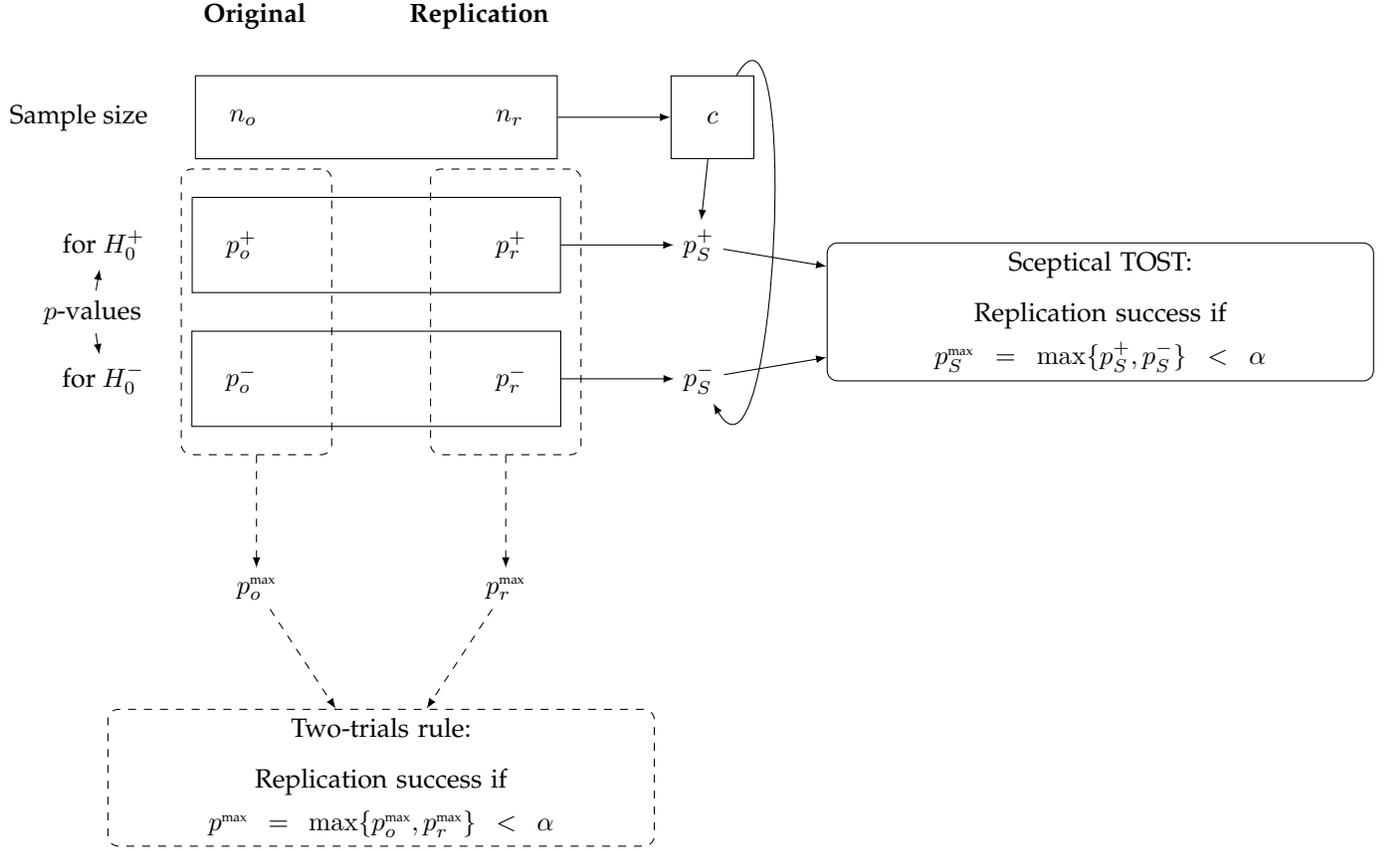
\begin{figure}[!h]
\centering
\vspace{1.5cm}
  \begin{tikzpicture}[node distance = 4.5cm, auto]
          \tikzset{
        box/.style = {
            fill=blue!15,
            shape=rectangle, 
            rounded corners,
                draw=blue!40, 
                align=center,
                text = black,
                font=\fontsize{12}{12}\selectfont},
        dummybox/.style = {
            shape=circle,  
                align=center, 
                minimum size={width("rrrrrrrrrrr")+2pt}},
        arrow/.style={
            color=black,
            draw=black,
            -latex,
                font=\fontsize{8}{8}\selectfont},
        }

\node (r1) [draw,inner sep=0.45cm, anchor = west]{ $p_o^{+}$ \hspace{3cm} $p_r^{+}$};
\node (r2) [below = 0.5cm of r1,draw,inner sep=0.45cm]{ $p_o^{-}$ \hspace{3cm} $p_r^{-}$};
\node (c) [above = 0.5cm of r1,draw,inner sep=0.45cm]{ $n_o$ \hspace{3cm} $n_r$};
\node (t1) [draw = none, left = 0.5cm of c]{Sample size};
\node (t2) [draw = none, left = 3cm of $(r1)!.5!(r2)$]{$p$-values};
\node (t3) [draw = none, left = 0.5cm of r1]{for $H_0^{+}$};
\node (t4) [draw = none, left = 0.5cm of r2]{for $H_0^{-}$};

  \node (original) [dashed, block, minimum width = 2cm, minimum height = 3.8cm, right = -2.6cm of  $(r1)!.5!(r2)$]{};
  \node (replication) [dashed, block, minimum width = 2cm, minimum height = 3.8cm, right = 1.3cm of original]{};
  
\node (orrepLabel)[above=0.5cm of c] {\textbf{Original \hspace{1.2cm} Replication}};

\node (cc) [draw, minimum height = 1.1cm, minimum width = 1.1cm, right = 1.5 cm of c]{$c$};
\node (pS2) [draw = none, right = 1.5 cm of r1]{$p_S^{+}$};
\node (pS1) [draw = none, right = 1.5 cm of r2]{$p_S^{-}$};

\node (pomax) [draw = none, below = 1.5 cm of original]{$\pomax$};
\node (prmax) [draw = none, below = 1.5 cm of replication]{$\prmax$};

\node (max) [block3, below = 1.6cm of  $(pomax)!.5!(prmax)$, dashed]{Two-trials rule: \\[.25cm] Replication success if \\[.15cm] $p^{\mbox{\tiny max}} = \max\{\pomax, \prmax \} < \alpha$};

\node (max2) [block3, right = 1.7cm of $(pS1)!.5!(pS2)$]{Sceptical TOST: \\[.25cm] Replication success if \\[.15cm] $ \pSmax = \max\{p_S^{+}, p_S^{-}\} < \alpha$};

% paths 

 \draw [arrow, dashed] (original) -- node [midway, right, dashed] {}(pomax);
  \draw [arrow, dashed] (replication) -- node [midway, right] {}(prmax);
  
  \draw[arrow, dashed] (pomax) -- node [midway, right, dashed] {} (max);
  \draw[arrow, dashed] (prmax) -- node [midway, right, dashed] {} (max);

   \draw [arrow] (c) -- node [midway, right] {}(cc);
  \path [arrow] (r1) -- node [midway, right] {}(pS2);
    \path [arrow] (r2) -- node [midway, right] {}(pS1);
    
     \path [arrow] (cc) -- node [midway, right] {}(pS2);
     \draw[arrow](cc) to [out=60,in=-60] node[below]{} (pS1);
     
       \draw[arrow] (pS1) -- node [midway, right] {} (max2);
  \draw[arrow] (pS2) -- node [midway, right] {} (max2);

     \draw[arrow] (t2) -- node [midway, right] {} (t3);
      \draw[arrow] (t2) -- node [midway, right] {} (t4);

  \end{tikzpicture}
 %  %
  \caption{Summary of the assessment of replication success for equivalence 
  studies with the 
  two-trials rule and the sceptical TOST procedure.}
  \label{fig:2tr_rs}
\end{figure}

We now adapt the sceptical $p$-value to the replication of equivalence studies. 
As there are two sets of hypotheses \eqref{eq:set2} and~\eqref{eq:set1}, 
the procedure explained above for superiority studies needs to be carried out twice,
with sceptical priors centered around $\delta$ and $-\delta$, 
respectively.
%
% For the hypotheses set~\eqref{eq:set2}, the original study is challenged 
% with a sceptical prior centered around $\delta$ and with 
% variance chosen such that the upper limit of the corresponding posterior 
% credible interval is equal to $\delta$.
The sceptical $p$-value $p_S^{+}$
measuring the conflict between the sceptical prior centered at $\delta$
and the replication 
data now depends on $p_o^{+}$, $p_r^{+}$ and $c$, see Figure~\ref{fig:2tr_rs}.
Similarly, 
% for the hypotheses set~\eqref{eq:set1}, 
% the original study is challenged 
% with another sceptical prior centered around $-\delta$ and with 
% variance chosen such that the lower limit of the corresponding posterior 
% credible interval is equal to $-\delta$. The resulting 
the sceptical $p$-value $p_S^{-}$
measuring the conflict between the sceptical prior centered at $-\delta$
and the replication 
data depends on $p_o^{-}$, $p_r^{-}$ and $c$.
Replication success is achieved if
$\pSmax = \max\{p_S^{+}, p_S^{-}\} <\alpha$.

%
%

% TEST 

% \begin{figure}[!h]
%   \centering

% \caption{Replication success with the sceptical TOST procedure.}
%   \label{fig:RS}
% \end{figure}

% ---------------------------------------------------------------------- %
% tikz

\subsection{Necessary conditions for replication success}\label{sec:upperBounds}
In superiority studies, the necessary replication success condition
 on the study-specific $p$-values is $\pmax < \alpha$
with the two-trials rule (which is also a sufficient condition)
and $\pmax < \gamma_c$ (which is not sufficient) with the sceptical $p$-value. The derivation of necessary bounds
is more 
involved for equivalence studies. We will use the same procedure for both methods.

First, the conditions $\pmax < \alpha$ and $\pSmax < \alpha$, respectively,
are rewritten as success intervals for $z_r^{+}$, which 
depend on $z_o^{+}$, $f_o$ and $c$.
We will then show that success is impossible
if $z_o^{+}$ is larger than a certain value.
% as the success interval 
% has a lower bound which is larger than the upper one.
%
This particular value of $z_o^{+}$ is transformed 
into a $p$-value $p_o^{+}$ which defines the largest value 
of $p_o^{+}$ for which replication success is possible.
The same procedure is then applied to obtain the necessary condition on $p_r^{+}$.
The details are provided in what follows.

% \todo[inline, caption={}]{
% \begin{itemize}
%   \item  monotonicity 
%   \item What does it mean if one line above other 
%   \item if one curve is smooth but not the other 
%   where is upper bound constant at 0.05.
% \end{itemize}
% }
% 
% \todo[inline]{CM: improve description of Figure 4, say why we do it, 
% sell it}
\paragraph{Two-trials rule}
The condition $\prmax < \alpha$ can be rewritten as a success 
region for $z_r^{+}$ conditional on the original study result:
\begin{eqnarray}\label{eq:succRegion_zrPlus}
z_\alpha - \frac{2 z_o^{+} \sqrt{c}}{f_o - 1} < z_r^{+} < - z_\alpha \, ,
\end{eqnarray}
see Appendix~\ref{sec:zr} for details.
However, if $z_o^{+} > z_\alpha(f_o - 1)/\sqrt{c}$, 
Equation \eqref{eq:succRegion_zrPlus} cannot be fulfilled as 
the lower bound for $z_r^{+}$ is larger than the upper one. Combining this 
with the condition $z_o^{+} < - z_\alpha$, the necessary condition on 
$z_o^{+}$ for replication success is 
\begin{eqnarray}\label{eq:bound_zoPlus}
z_o^{+} < z_\alpha\min\{(f_o - 1)/\sqrt{c}, - 1\} \, .
\end{eqnarray}
The right-hand side of~\eqref{eq:bound_zoPlus} is then transformed 
into an upper bound on the corresponding $p$-value $p_o^{+}$ for
which the two-trials rule can be fulfilled.
Note that this upper bound cannot be larger than $\alpha$.
Similarly, the condition $\pomax < \alpha$ can be rewritten 
as a success region for $z_o^{+}$, and the 
largest value of $p_r^{+}$ for which the two-trials rule can be 
fulfilled based on $f_r$ and $c$
can be derived, see Appendix~\ref{sec:zo} for detail.

% \hl{Figure~\ref{fig:upperboundpo} shows the upper bound 
% on $p_o^{+}$ (respectively on $p_r^{+}$) as a function of $f_o$ (respectively $f_r$).}
% \hl{The upper bound on $p_o^{+}$ and $p_r^{+}$ is by definition bounded by 
% $\alpha$, and is smaller than $\alpha$ with small values of $f_o$, 
% respectively $f_r$,
% for certain relative sample sizes $c$.}

\paragraph{Sceptical TOST procedure}
The replication success condition $\pSmax < \alpha$
can be expressed as a success region for $z_r^{+}$ conditional on the original 
study result, 
\begin{equation}\label{eq:succRegion}
\underbrace {z_{\gamma} \sqrt{1 + c \Big/ \left\{\left(\frac{f_o + 1}{f_o - 1}\right)^2K_o^{+} - 1\right\}} - \frac{2z_o^{+}\sqrt{c}}{(f_o-1)}}_{(z_r^{+})_{\mbox{\tiny lower}}} \leq z_r^{+}
\leq \underbrace{- z_{\gamma} \sqrt{\frac{1 + c}{(K_o^{+} -1)}}}_{(z_r^{+})_{\mbox{\tiny upper}}} \, ,
\end{equation}
with $K_o^{+} = (z_o^{+}/z_\gamma)^2$,  where 
$z_\gamma = \Phi^{-1}\left(1 - \gamma_c\right)$.
Details are given in Appendix~\ref{sec:zr}.
Success is thus impossible if $(z_r^{+})_{\mbox{\tiny lower}} >
(z_r^{+})_{\mbox{\tiny upper}}$, and root-finding algorithms can be used to calculate
the smallest value of $z_o^{+}$ where this happens.
This corresponds to an upper bound on $p_o^{+}$.
Equation~\eqref{eq:succRegion} can similarly be rewritten as a success region for 
$z_o^{+}$, and the upper bound on $p_r^{+}$ can be derived.

\begin{figure}[!h]
\begin{knitrout}
\definecolor{shadecolor}{rgb}{0.969, 0.969, 0.969}\color{fgcolor}
\includegraphics[width=\maxwidth]{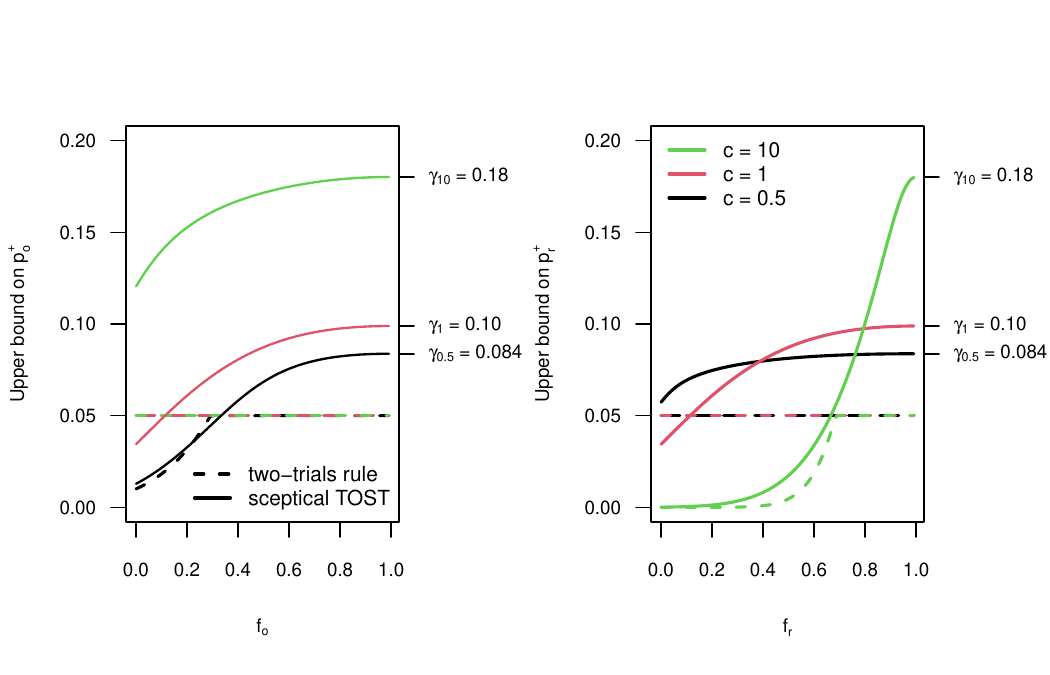} 
\end{knitrout}
\caption{Upper bound on $p_o^{+}$ (left) or $p_r^{+}$ (right) for replication success with the two-trials rule 
and the sceptical TOST procedure as a function of $f_o$ or $f_r$
for relative sample size $c \in \{0.5, 1, 10\}$ with $\alpha = 0.05$.}
\label{fig:upperboundpo}
\end{figure}

\paragraph{Comparison of the two methods}
Figure~\ref{fig:upperboundpo} shows the upper bound on $p_o^{+}$ (left)
and $p_r^{+}$ (right) with the two-trials rule and the sceptical 
TOST procedure as a function of $f_o$ and $f_r$, respectively.
Necessary conditions on 
$p_o^{-}$ and $p_r^{-}$ can be obtained directly
from the necessary conditions on $p_o^{+}$ and $p_r^{+}$
using Equation~\eqref{eq:zo2zo}, and are not shown.
With both methods, the upper bound on $p_i^{+}$, $i \in \{o, r\}$, 
increases as a function of $f_i$.
This makes sense, as the closer $f_i$ is to one, the smaller 
$p_i^{-}$ is (see Equation~\eqref{eq:zo2zo}) and so the more 
convincing the study is. The two-trials rule has the in-built condition
$p_i < \alpha$, explaining the plateau at $0.05$. The sceptical 
TOST procedure, on the other hand, tends 
to $\gamma_c > \alpha$ for $f_i \to 1$, as 
in the superiority case.
The influence of $c$ differs in the two plots. The upper bound 
on $p_o^{+}$ increases as a function of $c$, while it depends on 
$f_r$ for the upper bound on $p_r^{+}$.

To summarize, the necessary condition on the study specific $p$-values $p_o^{+}$
and $p_r^{+}$ is usually more stringent with the two-trials rule 
than with the sceptical TOST procedure. Non-significant studies can lead to replication success
with the latter method, but not with the former.

% The main difference between the two approaches can be seen in 
% Figure~\ref{fig:upperboundpo}:
% \hl{non-significant 
% studies ($\pomax > \alpha$ or $\prmax > \alpha$) can lead to replication success
% with the sceptical TOST method, but not with the two-trials rule. }
% Smaller $f_o$ (respectively $f_r$), lead to more stringent bounds on $p_o^{+}$
% (respectively $p_r^{+}$). 
% This makes sense, as 
% for the same $p_i^{+}$, a study with an effect estimate close to the
% margin ($f_i$ close to $1$) will have a smaller $p_i^{-}$
% than a study with an effect estimate close to 0 ($f_i$ close to $0$). 
% As $f_i$ tends to $1$, the upper bound on $p_i^{+}$ tends to $\gamma_c(\alpha)$
% as in the superiority setting.
% %
% Moreover, the upper bound on $p_o^{+}$ increases as the relative sample size $c$ 
% increases. In contrast, whether the upper bound on $p_r^{+}$ increases or 
% decreases as a function of $c$ depends on the value of $f_r$.

% here

\subsection{Application}\label{sec:appli}
The two methods are now applied to both examples presented
in Figure~\ref{fig:Goetz2011}. 
Table~\ref{tbl:pval} shows 
$\pmax$ from the two-trials rule and $\pSmax$ from the 
sceptical TOST procedure. In both examples, $\pmax$ and $\pSmax$ 
are similar in magnitude.
In the first study pair \citep{Goetz2011}, 
both methods lead to the opposite conclusion to that of the replication team: 
replication success cannot be declared at level $\alpha = 0.05$ as $\pmax = 
0.14$ and $\pSmax = 0.10$. 
However, at level $ 0.10 <\alpha < 0.14$, the sceptical 
TOST procedure would flag replication success
but not the two-trials rule, 
highlighting that significance of both studies is not
required with the former approach.
Replication success cannot be declared with either method in the second 
study pair \citep{Lin2012} regardless of the level $\alpha$ 
as the replication effect estimate $\hat\theta_r$ is larger than the margin
$\delta$
(in absolute value).

% \todo[inline]{CM: currently le text en vrac; come back and rewrite better +
% no replication success with either method+ too short at the moment}
% Table~\ref{tbl:pval} shows the sceptical TOST $p$-values for both 
% running examples presented in Section~\ref{sec:runningex}.
% The same conclusions are drawn as with the
% two-trials rule: replication success cannot be declared for either 
% of the examples.
% Table~\ref{tbl:pval} shows the TOST $p$-values of the running 
% examples \hl{with their respective margin}. 
% \hl{In both examples, the original effect estimate was approximately half 
% of the margin, $f_o = round(f_goetz, 2)$ and $f_o = - round(f_lin, 2)$, 
% respectively.}
% In the first study pair, the 
% two-trials rule leads to opposite conclusions to those of
% the respective replication teams.
% Equivalence cannot be declared
% at level $\alpha = alpha$
% in the original studies from \citet{Goetz2011} as
% $\pomax = p_o^{+} = biostatUZH::formatPval(study$po2[1])$.
% Equivalence can also not be declared in the second study pair as
% $\pomax = p_o^{+} = biostatUZH::formatPval(study$po2_strict[2])$ and
% $\prmax = p_r^{-}  = biostatUZH::formatPval(study$pr1_strict[2])$.
% The same conclusions be drawn from examining the 90\%-CI in
% Figure~\ref{fig:Goetz2011}.

% \todo[inline]{CM: say more about it, that success would be 
% achieved at 10\% but impossible in Lin because other direction etc}

% latex table generated in R 4.1.2 by xtable 1.8-4 package
% Thu Apr  4 17:21:54 2024
\begin{table}[ht]
\centering
\begin{tabular}{cccccccccc}
  \hline
   \multicolumn{4}{c}{} & \multicolumn{3}{c}{two-trials rule} & \multicolumn{3}{c}{sceptical TOST} \\  \cmidrule(lr){5-7} \cmidrule(lr){8-10}
  Study & $\hat\theta_o$ & $\hat\theta_r$ & $\delta$ & $p_o^{\mbox{\tiny max}}$ & $p_r^{\mbox{\tiny max}}$ & $p^{\mbox{\tiny max}}$ & $p_S^{-}$ & $p_S^{+}$ & $p_S^{\mbox{\tiny max}}$ \\ 
  \midrule
Goetz et al & 0.17 & 0.03 & 0.36 & 0.14 & 0.005 & {\bfseries 0.14} & 0.003 & 0.10 & {\bfseries 0.10} \\ 
  Lin et al & 0.09 & -0.21 & 0.18 & 0.062 & 0.78 & {\bfseries 0.78} & 0.84 & 0.016 & {\bfseries 0.84} \\ 
   \hline
  \end{tabular}
\caption{Original and replication effect estimates,
                  margin, two-trials rule 
                  and sceptical TOST 
                  $p$-values of the running examples.} 
\label{tbl:pval}
\end{table}

\section{Power and Type-I error rate}\label{sec:power}
Suppose now that neither of the two studies has been conducted yet.
Of great interest are the operating characteristics of the two methods. 
In particular, our focus lies in the probability to declare replication 
success under the alternative hypothesis that both effects are exactly $0$ 
(\emph{project power}), the intersection null that both effects are at 
the margin $\delta$ (\emph{overall T1E rate}), 
and under the union null that at least one effect is at the margin $\delta$ 
(\emph{partial T1E rate}).
The T1E rate varies with the value of $\theta \geq \delta$
and reaches a maximum for $\theta = \delta$ \citep[Section 11.5.2]{mat2006}, 
so the overall and partial T1E rates represent upper bounds.

\subsection{Project power}
% -------------
Let's denote the power of the original study to detect 
perfect equivalence ($\theta = 0$) by $1 - \beta$.
The power of a replication study of size $n_r = c\, n_o$ 
to reach a significant TOST procedure 
is then 2$\Phi\left(c \mu - z_\alpha \right) - 1$, 
with $\mu = z_{\beta/2} + z_\alpha$ \citep[Section 11.5.2]{mat2006}, 
where $z_{\beta/2} = \Phi^{-1}\left(1 - \beta/2\right)$.
The project power of the two-trials rule is the product of the two, 
\begin{eqnarray*}\label{eq:projPower_2tr}
(1 - \beta)\Phi\left(c \mu - z_\alpha - 1\right) \, ,
\end{eqnarray*}
as original and replication studies are independent.
The project power of the TOST procedure is the probability 
of \eqref{eq:succRegion} under the alternative hypothesis that the effect 
is exactly $0$ in both studies. The lower bound $(z_r^{+})_{\mbox{\tiny lower}}$ in~\eqref{eq:succRegion} 
depends on $f_o$ 
which can be rewritten as $f_o = z_o^{+}/\mu + 1$. Using the distributions $z_o^{+}\sim \Nor(- \mu, 1)$ and 
$z_r^{+} \sim \Nor(-\mu \sqrt{c}, 1)$
under the alternative hypothesis of perfect equivalence,
the project power is 
\begin{eqnarray*}\label{eq:projPower}
\int_{a}^{b}
\left[\Phi \left\{(z_r^{+})_{\mbox{\tiny upper}} + \mu \sqrt{c}\right\} -
\Phi \left\{(z_r^{+})_{\mbox{\tiny lower}} + \mu \sqrt{c} \right\} \right] \phi(z_o^{+} + \mu) d z_o^{+} \, ,
\end{eqnarray*}
where
\begin{eqnarray}\label{eq:ab}
a = z_{\gamma} - \mu \mbox{ and }
b = - z_{\gamma} + \mu \, .
\end{eqnarray}

Figure~\ref{fig:errRates} (a) shows the project power as 
a function of the relative sample size $c$ for an original power $1 - \beta$
of 80\%. The project power of the sceptical TOST procedure is always 
larger than of the two-trials rule, and is even larger than the original power 
for large values of the relative sample size $c$. 

\subsection{Overall T1E rate}
The T1E rate of the 
original study depends on $1-\beta$, the original power to detect 
perfect equivalence ($\theta_o = 0$). Namely, it is 
  \begin{eqnarray}\label{eq:oriPow}
  \alpha - \Phi(-2 z_{\beta/2} - z_\alpha) \, ,
  \end{eqnarray}
so it tends to $\alpha$ as $1 - \beta$ increases. 
Similarly, the replication T1E rate is 
  \begin{eqnarray}\label{eq:repliPow}
\alpha - \Phi\left(- 2 \sqrt{c} \mu + z_\alpha\right)
  \end{eqnarray}
  and depends both on the original power and the replication 
  sample size $n_r = c \,n_o$.
Because of independence of the two studies, the overall T1E rate of the two-trials rule
is the product of~\eqref{eq:oriPow} and~\eqref{eq:repliPow}.

Under the intersection null hypothesis that both effects are at the margin $\delta$, 
$z_o^{+} \sim \Nor(0, 1)$ and $z_r^{+} \sim \Nor(0, 1)$.
Using~\eqref{eq:succRegion}, the overall T1E rate of the sceptical TOST 
procedure is hence 
\begin{eqnarray*}\label{eq:overallT1E}
\int_{a}^{b}
\left[\Phi \left\{(z_r^{+})_{\mbox{\tiny upper}}\right\} -
\Phi \left\{(z_r^{+})_{\mbox{\tiny lower}} \right\} \right] \phi(z_o^{+}) d z_o^{+} \, ,
\end{eqnarray*}
with $a$ and $b$ defined in~\eqref{eq:ab}.
The overall T1E rate of both strategies is almost the same
(see Figure~\ref{fig:errRates} (b)): 
It starts at $0$ for small relative sample sizes $c$ and increases up 
to $\alpha^2 = 0.05^2 = 0.0025$.
This is in line 
with how the controlled sceptical $p$-value for superiority studies was 
constructed to match the overall T1E rate of the two-trials rule.
It is not constant at $\alpha^2$ as in the superiority setting 
as the replication power to detect perfect equivalence also matters here and is low 
for small relative sample sizes $c$.

\subsection{Partial T1E rate}
The partial T1E rate is the probability of a false claim of replication success
with respect to the union null hypothesis. We assume here that there is perfect
equivalence in the first study ($\theta_o = 0$) and no equivalence in the second 
study ($\theta_r = \delta$). 
The partial T1E rate of the two-trials rule is the product of the original power
$1-\beta$ and the replication T1E rate expressed in~\eqref{eq:repliPow}, 
because of independence of the two studies.

\begin{figure}[!h]
  \centering
\begin{knitrout}
\definecolor{shadecolor}{rgb}{0.969, 0.969, 0.969}\color{fgcolor}
\includegraphics[width=\maxwidth]{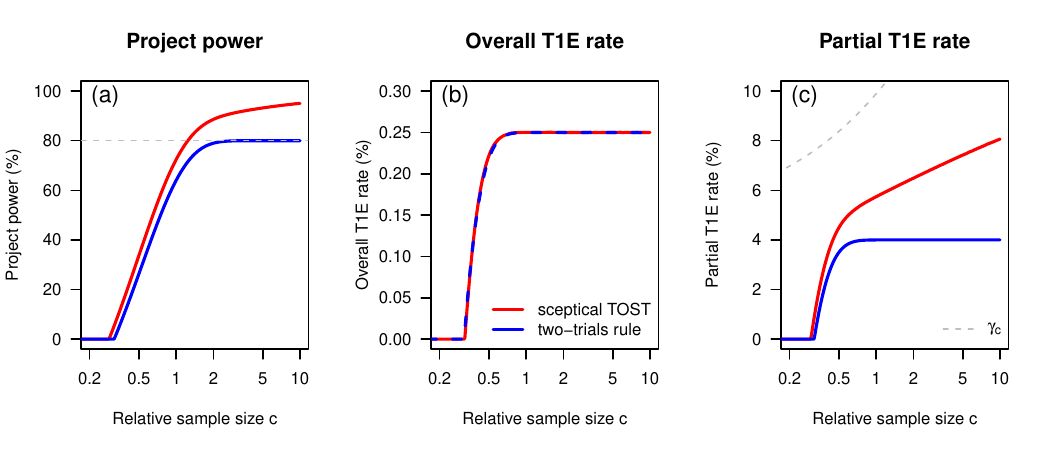} 
\end{knitrout}
\caption{Project power, overall T1E rate,  and partial T1E rate of the two-trials rule and
the sceptical TOST procedure.
The original study has a power of $80$\%  to detect 
perfect equivalence ($\theta_o = 0)$.}
\label{fig:errRates}
\end{figure}

The distributions 
of $z_o^{+}$ and $z_r^{+}$ are $z_o^{+} \sim \Nor(\mu, 1)$ and $z_r^{+} \sim \Nor(0, 1)$.
Using~\eqref{eq:succRegion}, the partial T1E rate of the sceptical TOST procedure is hence 
\begin{eqnarray*}\label{eq:partialT1E}
\int_{a}^{b} 
\left[\Phi \left\{(z_r^{+})_{\mbox{\tiny upper}}\right\} - 
\Phi\left\{(z_r^{+})_{\mbox{\tiny lower}}\right\} \right]\phi(z_o^{+} + \mu) d z_o^{+} \, .
\end{eqnarray*}
The partial T1E rate of both methods is shown in Figure~\ref{fig:errRates} (c).
For the same reason as explained in the previous section, 
it is $0$ for small relative sample sizes $c$.
Moreover, the partial T1E rate of the two-trials is constant at $0.05 \times 0.8 = 4$\%
if the relative sample size $c$ is large enough. 
In contrast, the T1E of the sceptical TOST procedure increases with $c$, 
and does not reach a plateau but
is always smaller than $\gamma_c$.
In a sense, the increase in partial T1E rate is the price to pay for the 
increase in project power of the sceptical TOST procedure. However, in practice the relative 
sample size $c$ is not fixed but chosen based on the original 
study. As a result, there is another T1E rate
that is more suited than the partial T1E rate: the conditional T1E rate 
described in Section~\ref{sec:condt1e}.

% new plots 

\section{Design of replication studies}\label{sec:design}
The previous section assumes that none of the equivalence
studies have been conducted yet. 
Here we assume that the results of the original study are available, and use them 
to calculate the sample size of the replication. 
As the design of a replication study should always match the planned analysis 
method \citep{Anderson2022}, we provide sample size calculations for 
the two-trials rule and the sceptical TOST procedure.

  \subsection{Sample size calculation}\label{sec:ss_calculation}
  % \todo[inline]{CM: Add standard references for samlpe size calculation for 
  % equivalence studies in particular Flight for equation (16).}
  %
  In superiority studies, the replication sample size is usually chosen such that the conditional 
  power to detect the original effect estimate $\hat\theta_o$ at level $\alpha$ 
  reaches a certain value (typically between 80 and 95\%). 
  This is the approach which has been taken 
  in the Reproducibility Project: Cancer Biology \citep{Errington2021}
  for the replication of significant original studies.
  However, this approach does not work for the original studies 
  with a null result as $\hat\theta_o$ is very close to $0$. For 
  these studies, the replication teams selected the sample size 
  based on a sensitivity analysis, see the corresponding 
  registered reports for more information \citep{Fiering2015, Blum2015}.
  This demonstrates the need for an appropriate sample size method
  targeted towards equivalence studies, 
  as explained in this section.
  We assume here that the true effect $\theta_o = \theta_r = \theta$
  is the same in both studies.

\paragraph{Two-trials rule}

The replication power of the TOST procedure can be calculated using the 
confidence interval perspective. 
Namely, it is the probability that the estimate $\hat\theta_r$ lies 
within $[- \delta + z_{\alpha}\sigma_r; \delta - z_{\alpha} \sigma_r]$.
Now $\hat\theta_r$ is $\Nor(\theta, \sigma_r)$ distributed, so the power
of the replication study at significance level $\alpha$ is
% to detect $\theta$ is 
%
\begin{eqnarray*}
\Phi \left(\frac{\delta - \theta}{\sigma_r} - z_\alpha \right) - 
\Phi \left(\frac{- \delta - \theta}{\sigma_r} + z_\alpha\right) \, .
\end{eqnarray*}
This corresponds to the standard power formula for equivalence 
studies \citep[][Equation (5)]{Flight2015}.
If the true effect size $\theta$ is assumed to be exactly $0$, the 
power simplifies to $2\Phi(\delta/\sigma_r - z_\alpha) - 1$. 
However, the original effect estimate $\hat\theta_o$ is a much more 
suited replacement for $\theta$. In our approach, we want to express 
the power on the relative scale and this gives the conditional power: 
% to detect $\hat\theta_o$:
%  
\begin{eqnarray}\label{eq:condpow.sign}
\CPTR & = \Phi (\underbrace{-z_o^{+} \sqrt{c} - z_{\alpha}}_{A})
- \Phi (\underbrace{-z_o^{-}\sqrt{c} + z_{\alpha}}_{B}) \, .
\end{eqnarray}
This approach assumes that the effect $\hat\theta_o$
detected in the original
study is the unknown, true effect $\theta$ and ignores
its inherent uncertainty.
Uncertainty of the original effect estimate $\hat\theta_o$ is
acknowledged in the power calculation
if the predictive distribution
\begin{eqnarray}\label{eq:d.distr_pred}
\hat\theta_r \sim \Nor \left(\hat\theta_o, \sigma_o^2 + \sigma_r^2 \right)
\end{eqnarray}
is used. 
This leads to the predictive power,
\begin{eqnarray}\label{eq:predpow.sign}
% \PPTR & = \Phi \left(\mbox{\it A}/\sqrt{c+1}\right)
% - \Phi \left(\mbox{\it B}/\sqrt{c+1}\right) \, .
\PPTR = \Phi\left(\frac{\mbox{\it A}}{\sqrt{c + 1}}\right)
- \Phi\left(\frac{\mbox{\it B}}{\sqrt{c + 1}}\right) .
\end{eqnarray}
Predictive power thus is not conditional on a single
value for $\theta$, but instead uses a distribution
of potential $\theta$ values provided by the original study.
The concept of predictive power has first been proposed 
by \citet{SpiegFreed1986} to provide a more direct predictive interpretation
than conditional power.
In the context of replication studies it
has been discussed 
in~\citet{MicheloudHeld2021} and \citet{HeldMichPaw2021}
for superiority designs. 
Its use has also been investigated for
(bio)equivalence studies
\citep[Section 2.1.4]{OHagan2005, Ring2019}.

\paragraph{Sceptical TOST procedure}

Equation~\eqref{eq:succRegion} can also be used to derive the
conditional power with the sceptical TOST procedure. The only difference 
as compared to the project power
is that $z_o^{+}$ and $f_o$ are now fixed as the results of the original study are known. 
The conditional power is then the probability of~\eqref{eq:succRegion}, namely
\begin{eqnarray}\label{eq:condpow.RS}
\CPRS = \Phi(\underbrace{-z_o^{+}\sqrt{c} - z_{\gamma}\sqrt{1 + c/(K_o^{+} - 1)}}_{C})
- \Phi(\underbrace{-z_o^{-}\sqrt{c} + z_{\gamma}\sqrt{1 + c/(K_o^{-} - 1)}}_{D}) \, .
\end{eqnarray}
The predictive power turns out to be 
\begin{eqnarray}\label{eq:predpow.RS}
\PPRS = \Phi\left(\frac{\mbox{\it C}}{\sqrt{c + 1}}\right)
- \Phi\left(\frac{\mbox{\it D}}{\sqrt{c + 1}}\right) .
\end{eqnarray}

The four power formulas \eqref{eq:condpow.sign}, \eqref{eq:predpow.sign},
\eqref{eq:condpow.RS}, \eqref{eq:predpow.RS} all contain two terms.
Due to these two components, the required relative sample size $c$ 
to reach a fixed power in the equivalence setting
cannot be derived by simply inverting the power formulas as in the superiority
framework. Instead, application of root-finding algorithms is necessary.

\subsection{Properties}
Figure~\ref{fig:cond_t1e} compares the required relative sample size $c$ 
to reach a conditional and predictive power of 80\%
with the two-trials rule and the sceptical TOST procedure.
Two cases are explored: $f_o = 0.1$, 
meaning the original effect estimate $\hat\theta_o$ was close to $0$, and 
$f_o = 0.9$, meaning that $\hat\theta_o$ 
was close to the margin $\delta$. 
With both methods, the required relative sample $c$ size is slightly smaller for larger 
$f_o$ as for a fixed $p_o^{+}$, $p_o^{-}$ decreases for increasing 
$f_o$.
Furthermore, the required sample size is smaller with the sceptical 
TOST procedure as compared to the two-trials rule for 
more convincing original studies, and larger for original studies 
with a $p$-value $p_o^{+}$ close to $\alpha = 0.05$. 
But the main difference between the two approaches is that, while 
$\pomax < 0.05$ is required in two-trials rule, there is 
no such dichotomization in the sceptical TOST procedure, where
sample size calculation is also possible based on 
non-significant original studies.
With conditional power, it is possible as long as $\hat\theta_o < \delta$, 
\ie $p_o^{+} < 0.5$, while it depends on the values 
$f_o$ and the target power with predictive power. For a predictive power 
of 80\%, replication sample size calculation is possible if $p_o^{+} < 0.13$ for 
$f_o = 0.1$ and if $p_o^{+} < 0.21$ for $f_o = 0.9$.
As in superiority studies, 
predictive power is typically smaller than conditional power with
both methods, hence 
a larger replication sample size is required if the calculation is 
based on the former \citep{MicheloudHeld2021}.

Note that
the upper bound on $p_o^{+}$ and $p_r^{+}$ with 
the sceptical TOST procedure is at most $\gamma_c$ 
(see Figure~\ref{fig:upperboundpo}), which 
increases with $c$. 
This means that a large relative sample size $c$ will make 
replication success in theory 
possible even if one of the $p$-values is very large. 
This might seem controversial,
however, we will show in Section~\ref{sec:condt1e} that the T1E 
rate of the replication study, conditional on the original study results, 
gets smaller for larger $p_o^{+}$, and so for larger relative sample sizes $c$. 
This point will also be further discussed in Section~\ref{sec:discussion}.

\subsection{Sample size recalculation}

% 
% In the replication study of \citet{Goetz2011} and \citet{Lin2012},
% the relative sample size $c$ was chosen as in superiority studies. 
% Namely, a relative sample size of $c = round(study$c[1], 1)$
% and $c = round(study$c[2], 1)$, respectively, had a replication 
% power of 80\% to detect a standardized mean difference of 
% $d = 1.4$ (Fisher's $z$ = $round(used_d_1, 2)$) and 
% $d = 0.14$ (Fisher's $z$ = $round(used_d_2, 2)$), respectively.

% Table~\ref{tbl:ss} shows the required relative sample size using a proper 
% equivalence methodology. No replication sample size can be calculated with the two-trials 
% rule at level $\alpha = 5\%$ as $p_o^{+} > 5\%$ in both original studies. This is not the case
% with the sceptical TOST procedure which always allows sample size calculation 
% even with non-significant original studies.
% However, the required relative sample size might be very high, 
% for example the replication study of \citet{Goetz2011}
% would required a relative sample size $c = 21.1$.

%

Using the appropriate equivalence methods developed above, 
the required relative sample size $c$ has been recalculated 
based on the original studies by \citet{Goetz2011} and \citet{Lin2012}.
For a conditional power of 80\% and $\alpha = 0.05$,
a relative sample size of $c = 21.1$
is required with the sceptical TOST procedure in the former study, and $c = 4.2$
in the latter. In both cases, the required relative sample size is larger 
than the one which was actually used in the two replication studies ($c = 1.9$, and 
$c = 2.1$, respectively).
No replication sample size can be calculated with the two-trials rule 
as $p_o^{+} > 5\%$ in both original studies.

\subsection{Conditional T1E rate}\label{sec:condt1e}
It is also possible to calculate the conditional Type-I error rate, 
\ie the T1E rate of the replication study, 
conditional on the original study result \citep{Micheloud2023}.
To do so, the relative sample size $c$ to reach a certain power based on a 
particular original result ($p_o^{+}$, $f_o$)
is calculated using the methodology 
presented in Section~\ref{sec:ss_calculation}.
The probability of \eqref{eq:succRegion} with the relative sample size $c$ 
and the null distribution $z_r^{+} \sim \Nor(\mu \sqrt{c}, 1)$ then gives 
the conditional T1E rate.

% Calculation of maximum cond t1e rate for other values of the power 
% 90 and 95

\begin{figure}[!h]
\centering
\begin{knitrout}
\definecolor{shadecolor}{rgb}{0.969, 0.969, 0.969}\color{fgcolor}
\includegraphics[width=\maxwidth]{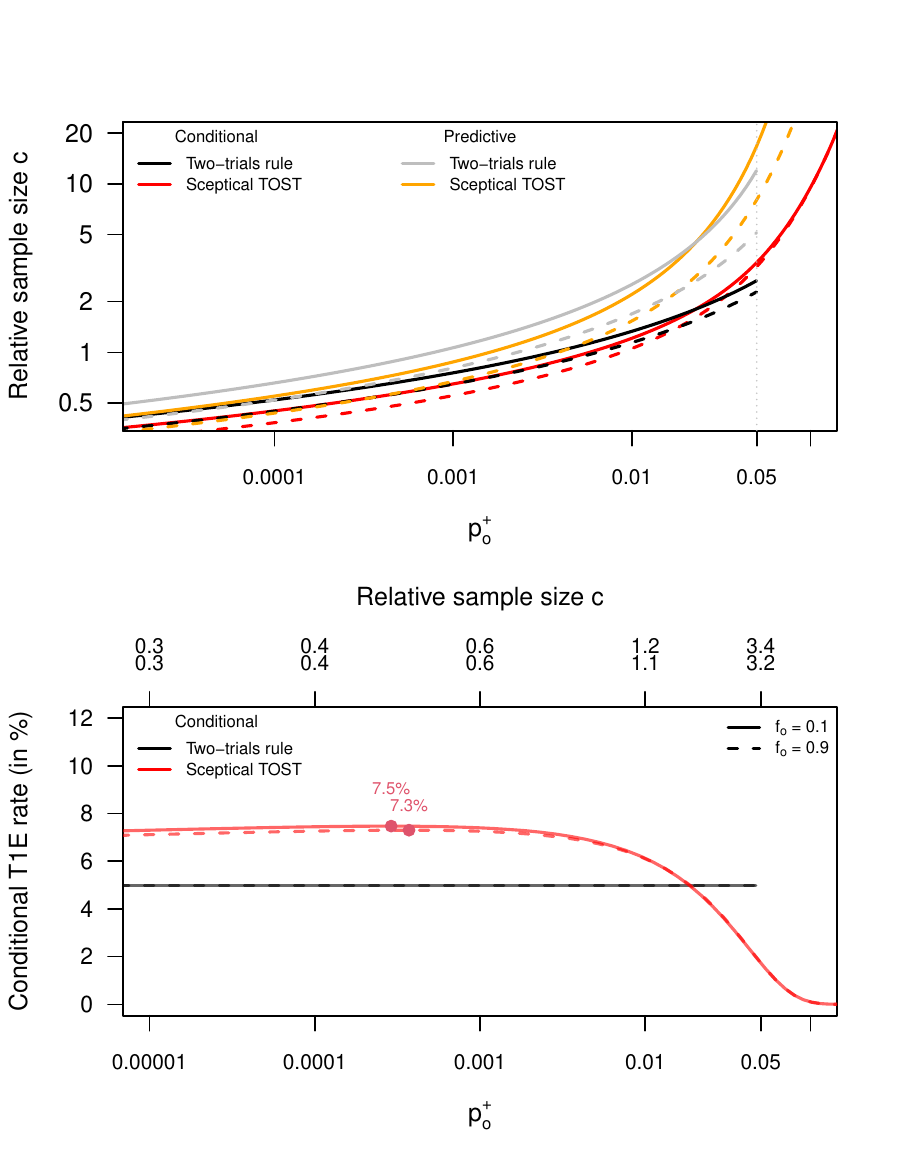} 
\end{knitrout}
\caption{Relative sample size $c$ to reach a conditional or predictive power of 80\% (top)
and conditional T1E with a conditional power of 80\% (bottom)
of the sceptical TOST procedure 
and the two-trials rule as a function of $p_o^{+}$ for $f_o = 0.1$ 
and $f_o = 0.9$.
The top axis in the bottom plot
shows the required relative sample size $c$ with $f_o = 0.1$ (first row) and $f_o = 0.9$ (second row)
to reach a conditional power of 80\%.}
\label{fig:cond_t1e}
\end{figure}

Figure~\ref{fig:cond_t1e} (bottom) shows the conditional T1E rate of both methods 
as a function of $p_o^{+}$. The relative sample size $c$ 
shown on the top axis is calculated 
using the sceptical TOST method to reach a conditional power of 
$80$\% and is very similar 
for  $f_o = 0.1$ (first row) and $f_o = 0.9$ (second row).
Regardless of $f_o$, the conditional T1E rate of the two-trials rule 
is fixed at $\alpha = 5\%$, provided that $p_o^{+} < 5\%$. In contrast, 
the conditional T1E rate of the sceptical TOST procedure 
is smaller than 5\% if $p_o^{+} < 0.019$,
and larger otherwise. However, it does not exceed
$7.5$\% (for $f_o = 0.1$) and $7.3$\% (for $f_o = 0.9$) for a conditional 
power of 80\%. For a conditional power of 90\% and 95\%, these maximum 
values are 7.8\% ($f_o = 0.1$) and
7.7\% ($f_o = 0.9)$, 
and 8\% (both $f_o = 0.1$ and $f_o = 0.9$), respectively.
It is
thus always smaller than 2$\alpha$ as in the superiority setting 
\citep[Section 3.4]{Micheloud2023}.

\section{Discussion}\label{sec:discussion}

Replication studies are becoming more and more popular and encouraged
in many domains of science.
However, the existing literature greatly focuses on superiority studies,
and non-significant studies are often incorrectly
interpreted as showing evidence for a null effect.
In order to bridge this gap,
we proposed two methods for the analysis and the design of 
replication studies with an equivalence design: 
the two-trials rule and the sceptical TOST procedure.

% The main difference between the two approaches is that the sceptical TOST 
% procedure does not have a strict dichotomization at $\alpha$ for the original 
% and replication TOST $p$-values.  
Both approaches exactly control the 
overall T1E rate at level $\alpha^2$ if the replication sample size is 
sufficiently large.
The main difference between the two approaches is that replication success 
is possible with the sceptical TOST 
procedure if one of the $p$-values is non-significant, but not with the 
two-trials rule. 
As a result, the sceptical TOST procedure has a larger project power 
than the two-trials rule. This increase in project power comes at the 
price of an increased partial T1E rate. 
However, the conditional T1E rate of the sceptical TOST procedure is 
smaller than the one of the two-trials rule for larger values of $p_o^{+}$, and is 
at most less than two times larger otherwise, which is considered 
acceptable by some authors \citep{Rosenkrantz2002}.
In addition, sample size calculation is possible
even for original studies with $\pomax > \alpha$ with the sceptical TOST 
procedure.
However, one might argue that the bound $\gamma_c$
on the partial T1E rate becomes unreasonably large
if $c$ is very large. To mitigate this issue, 
an alternative would be to use a method with partial T1E rate not depending on $c$, 
such as Edgington's method
\citep{Held2024arxiv}. 

Some limitations need to be noted. 
Our methodology relies on normality of the effect estimates, 
which might seem restrictive. However
many framework fall into this category after a suitable transformation. 
% Moreover, the margins are assumed to be symmetric around $0$, 
% an aspect which could be generalized in a subsequent paper.
It also needs to be noted that two study pairs from \citet{Errington2021} are 
for illustrative purposes only, and the results depend
on the choice of the margin. A sensitivity analysis could be
performed by looking at $\pmax$ and $\pSmax$
as a function of the margin \citep{Hauck1986}.

Future work will focus on developing a sceptical confidence interval
using the $p$-value function \citep{Infanger2019}. Replication success 
could then be assessed by verifying whether the sceptical confidence interval is 
contained within the equivalence region.

\section*{Data and software availability}

The dataset used in this paper has been downloaded from 
\url{https://osf.io/39s7j} \citep{dataRPCB}.
The code to reproduce all the analyses and figures can be 
found at \url{https://gitlab.uzh.ch/charlotte.micheloud/repromat-equivalence}, 
together with the R-package \texttt{RepEquivalence} which contains 
functions for the design and analysis of replication studies with 
an equivalence design.

\section*{Acknowledgments}
Support by the Swiss National Science Foundation (Project \# 189295)
is gratefully acknowledged. We thank the reviewer and the associate 
editor for their helpful comments.

\section*{Conflict of interest}
The authors have no conflict of interest to declare.

\vspace{2cm}
% \begin{center}

{\LARGE \bf Appendix}
% \end{center}
\appendix

\section{Success region for $z_r^{+}$}\label{sec:zr}
\paragraph{Two-trials rule}
The condition on $z_r^{+}$ for replication success is $z_r^{+} < - z_\alpha$ ,
and the condition on $z_r^{-}$ is 
$z_r^{-} > z_\alpha$ .
Moreover, $z_r^{-}$ can also be expressed as 
\begin{eqnarray}\label{eq:zr_complicato}
z_r^{-} = (\hat\theta_r + \delta) /\sigma_r = z_r^{+} + \frac{2 z_o^{+} \sqrt{c}}{f_o - 1} \, ,
\end{eqnarray}
so the success region for $z_r^{+}$ is 
\begin{eqnarray*}
z_\alpha - \frac{2 z_o^{+} \sqrt{c}}{f_o - 1} < z_r^{+} < - z_\alpha
\end{eqnarray*}

\paragraph{Sceptical TOST procedure}
Following \citet[Equation (16)]{Micheloud2023} with $z_o = z_o^{+}$
and $z_r = z_r^{+}$, the condition on $z_r^{+}$ for replication success with the 
sceptical TOST procedure is
\begin{eqnarray}\label{eq:zrPlus}
z_r^{+} < - z_\gamma \sqrt{1 + c/(K_o^{+} - 1)} \, ,
\end{eqnarray}
and similarly, 
the condition on $z_r^{-}$ is
\begin{eqnarray}\label{eq:zrMinus}
z_r^{-} > z_\gamma \sqrt{1 + c/(K_o^{-} - 1)} \, .
\end{eqnarray}
Now, using \eqref{eq:zo2zo} and \eqref{eq:zr_complicato}, 
Equation~\eqref{eq:zrMinus} can be rewritten as 
\begin{eqnarray}\label{eq:zrPlus_bis}
z_r^{+} > z_\gamma \sqrt{1 + c/\left\{\frac{(f_o + 1)^2}{(f_o - 1)^2}K_o^{+} - 1\right\} } - 
\frac{2z_o^{+}\sqrt{c}}{f_o - 1} \, .
\end{eqnarray}
Putting~\eqref{eq:zrPlus} and ~\eqref{eq:zrPlus_bis} together 
gives the success region for $z_r^{+}$.

\section{Success region for $z_o^{+}$}\label{sec:zo}

\paragraph{Two-trials rule}
The condition on $z_o^{+}$ for replication success is
% \begin{eqnarray*}
$z_o^{+} < - z_\alpha$ ,
% \end{eqnarray*}
%
and the condition on $z_r^{-}$ is 
% \begin{eqnarray*}
$z_o^{-} > z_\alpha$.
% \end{eqnarray*}
%
Moreover, $z_o^{-}$ can also be expressed as 
\begin{eqnarray}\label{eq:zo_complicato}
z_o^{-} = \frac{\hat\theta_o + \delta}{\sigma_o} = z_o^{+} + \frac{2 z_r^{+}}{\sqrt{c}(f_r - 1)} \, ,
\end{eqnarray}
so the success region for $z_o^{+}$ is 
\begin{eqnarray*}
z_\alpha - \frac{2 z_r^{+}}{(f_r - 1)\sqrt{c}} < z_o^{+} < - z_\alpha \, .
\end{eqnarray*}

\paragraph{Sceptical TOST procedure}
Equations~\eqref{eq:zrPlus} and ~\eqref{eq:zrMinus} are rearranged to 
give a condition on $z_o^{-}$ and $z_o^{+}$, respectively. 
Using~\eqref{eq:zo2zo} and \eqref{eq:zo_complicato}, the success region of 
$z_o^{+}$ is 
\begin{eqnarray*}
z_\gamma \sqrt{1 + \frac{c}{(K_r^{-} - 1)}} - \frac{2 z_r^{+}}{(f_r - 1)\sqrt{c}}
< z_o^{+} < - z_\gamma \sqrt{1 + \frac{c}{(K_r^{+} - 1)}} \, .
\end{eqnarray*}
\bibliographystyle{apalike} 
\bibliography{biblio}
\end{document}